\documentclass[twocolumn,pra,aps]{revtex4}
\usepackage{dcolumn}
\usepackage{amsmath,amssymb}
\usepackage{boxedminipage}
\usepackage{array}
\usepackage[mathscr]{euscript}
\usepackage{verbatim}
\usepackage{graphicx}
\usepackage{epsfig}
\usepackage{epstopdf}
\newcolumntype{.}{D{.}{.}{1}}
\begin{document}
\title{Structural changes in quasi- 1D many-electron systems: from linear to zig-zag and beyond}
\author{R. Cortes-Huerto, M. Paternostro, and P. Ballone}
\affiliation{School of Mathematics and Physics, Queen's University Belfast,
Belfast BT7 1NN, UK}
\date{\today}
\begin{abstract}
Many-electron systems confined to a quasi-1D geometry by a cylindrical distribution of positive charge have been 
investigated by density functional computations in the unrestricted local spin density approximation. Our investigations have been
focused on the low density regime, in which electrons are localised. The results reveal a wide variety of different charge and spin configurations,
including linear and zig-zag chains, single and double-strand helices, and twisted chains of dimers. The spin-spin coupling turns from weakly 
anti-ferromagnetic at relatively high density, to weakly ferromagnetic at the lowest densities considered in our computations. 
The stability of linear chains of localised charge has been investigated by analysing the radial dependence of the self-consistent potential and by 
computing the dispersion relation of low-energy harmonic excitations.
\end{abstract}
\maketitle
\newpage

\section{Introduction}

Systems made of ionised atoms confined into Penning~\cite{penning} or Paul~\cite{paul} traps provide interesting prototypes of low-dimensional many-particle systems.
Confinement is enforced by the application of static (Penning) and/or time-dependent electromagnetic fields (Paul) and temperature can be reduced to 
the mK range by laser cooling~\cite{laser}. Systems composed of atoms ranging in number from  a few tens to several thousand are routinely made and their 
properties analysed by a broad range of spectroscopic techniques~\cite{rmp}.

Fine control of the ion density $\rho_I$ and temperature $T$ allows to experimentally probe a wide range of inter-particle couplings, whose strength is measured 
by the ratio $\Gamma$ between the potential and kinetic energy of the ions. At sufficiently high coupling ($\Gamma \sim 180$, see Ref.~[\onlinecite{baus}]), ions condense
into a regular lattice, fulfilling early predictions generally attributed to Wigner~\cite{wigner}. Such a transition has been observed
in trapped clouds made of single~\cite{single} or binary~\cite{mixtures} ion species, providing an intriguing view of ordered Coulomb systems at low 
temperature and high couplings~\cite{nat, win}. Upon changing $\rho_I$, $T$ or the shape and strength of the applied fields, these systems 
undergo a sequence of characteristic structural changes such as order-disorder~\cite{diedrich} and isomerisation transitions~\cite{isomers}. 

Recently, experiments~\cite{birkl, waki} and computational studies~\cite{fishman} have focused on one specific phase change taking
place in nearly 1D trapped-ion systems, transforming linear chains into zig-zag configurations~\cite{report}.  Simple consideration of the forces
active in and on the system suggests that such a transition arises from the competition between the interaction with the external field
(increasing upon the transition) and the  electron-electron repulsion, which decreases because of the larger nearest neighbour distance 
in zig-zag chains. The order of the transition is not precisely known and, strictly speaking, not even well defined for the finite samples
probed in experiments. However, computations for extended systems with periodic boundary conditions suggest that the transition is continuous and second order, with 
a discontinuity in the second derivative of the ground state energy with respect to the 1D ion density~\cite{fishman}.

The linear to zig-zag transition has a number of different implications on the system properties. In the case of ions carrying a magnetic (spin) moment, 
for instance, changing the amplitude of the zig-zag modulation changes the number and relative distance of
the ions' neighbours, and, by varying the relative weight of first and second nearest neighbour interactions, it provides a way to tune 
the spin-spin coupling~\cite{kli}. At the same time, the zig-zag transition doubles the unit cell of the system, thus changing even qualitatively the vibrational spectrum of 
the chain. Interestingly, the dimerisation that often accompanies the zig-zag transition could open the way to the Bose-Einstein condensation of 
ions whose spin is half integer. Finally, the link between the structural transitions seen in experiments and the behaviour of quantum correlations among the
trapped ions has been established by a theoretical study based on techniques typical of continuous-variable systems~\cite{vedral}. 

Besides providing an appealing playground to investigate the interplay between dimensionality and many-particle effects, quasi-1D atomic plasmas are actively 
investigated in view of applications in metrology~\cite{metrol} and in quantum information technology~\cite{cirac}.
Moreover, low temperature ions confined in a quasi-1D ion trap have been proposed as models for the analogic simulation of many-particle systems~\cite{porras, fried, simul}.
In most of these applications, quantum mechanics plays an important role. This observation has motivated us to study quasi-1D systems made
of electrons, whose light mass amplifies the quantum mechanical effects. To the best of our knowledge, low-dimensional many-electron plasmas confined at low
temperature into electromagnetic traps have not been made and characterised in experiments. However, it should be mentioned that there is a considerable interest 
in scaling up the already experimentally demonstrated ability to trap and coherently control a single electron in a Penning trap to a genuine quantum many-body 
configuration. Roadmaps towards the achievement of such a situation have been detailed in Ref.~[\onlinecite{electrontrap}].
It is worth reminding that closely related systems, consisting of mobile electrons in conducting nano-wires can be prepared by a variety of methods, including the 
controlled doping of semiconducting nanostructures~\cite{pepper, Werner06} and conducting polymers~\cite{wig, Rahman06, roth}. Moreover, new fabrication methods 
are being developed based on the doping or the electrostatic biasing of carbon nanotubes~\cite{carbon}.

The properties of electrons in these systems are often described using the one-component plasma picture~\cite{zabala}, thus representing electrons as independent particles 
moving in a fixed external potential. We resort to the simplest version of this approach, based on the jellium model~\cite{vignale}, in which the external potential 
confining the electrons is due to their Coulomb interaction with a cylindrical background of positive charge, whose density is constant ($\rho_b$) 
within a pre-defined volume and zero outside. In order to approach the conditions of interest for charged particles in a trap, we consider the limit of very thin 
wires with a large aspect ratio between the length $L_z$ and the radius $R_b$  of the background charge distribution ($ 27.7 \leq L_z / R_b \leq 39.2$). In our computation, many-body effects are accounted for by resorting to the simple local-spin density (LSD) approximation~\cite{PZ} to density functional (DF) 
theory~\cite{ks}.  Single electron orbitals are expanded on a large basis of plane waves, and the ground state energy and density are determined by
direct minimisation, without any symmetry restrictions. We focus our attention on the low density, high correlation regime, where the Wigner crystal is the stable phase, 
and we consider various combinations of spin-up and spin-down populations.

Our calculations demonstrate the existence of linear and zig-zag chains, stabilised by different combinations of  the 1D electron density, spin configuration
and shape of the confining potential. Furthermore, our results display other unforseen structures, never considered or found so far for classical ion systems. 
Dimers, already suggested long ago for quantum spin chains~\cite{gosh}, appear at low density. At intermediate densities, we find new geometries, such as helices, and even 
double helices. Hints of these exotic geometries were already given by calculations for larger wires at much lower density~\cite{hughes}. 

In addition to this basic information on the ground state density and spin configuration, our results provide a wealth of new data on the density of
states, the electric conductivity and the vibrational modes of nearly 1D electron systems. In particular, we give strong evidence of a second-order 
nature of the linear-to-zig-zag transition, a result that appears to be in line with the findings in Ref. [15].

The paper is organised as follows. The model and the computational method are defined and briefly discussed in Section~\ref{method}. The computational
results for the ground state density and spin distribution are described in Sec.~\ref{ground}, while the computation of phonon-like excitations
is reported in Sec.~\ref{phons}. A summary and a brief outline of promising new directions are given in Sec.~\ref{summary}.

\section{model and methods}
\label{method}
Computations have been carried out for systems of $N=N_{up}+N_{dn}$ electrons, neutralised by a cylindrical background of positive charge, whose axis is
parallel to the $z$ direction. In what follows, the background density $\rho_b$ is expressed in terms of the Wigner-Seitz radius $r_s$ through 
the relation $\rho_b=3/4\pi r_s^3$. Here $N_{up}$ and $N_{dn}$ are the number of spin-up and spin-down electrons, respectively. 
The length $L_z$ and radius $R_b$ of the cylindrical background satisfy the neutrality condition $\pi R_b^2 L_z \rho_b=N$. 
Moreover, the number of electrons per unit length of the wire is $\rho_{lin}=\pi R_b^2 \rho_b$.  Atomic units are used throughout the paper, and cylindrical
coordinates $(r, \phi, z)$ are implicitly assumed in our equations and description of the results.

The basic cylindrical segment described above is periodically 
replicated in the direction parallel to the $z$ axis with periodicity $L_z$, thus representing an extended wire along such direction.
Due to our choice of plane waves as basis functions (see below), we periodically replicated our sample also in the $xy$ plane.
For the sake of simplicity, we adopt the same periodicity $L_z$ in all three directions.

The ground state energy and density are computed within the Kohn-Sham (KS) formulation of density functional theory (DFT), in which electrons occupy 
single-particle KS states $\{ \psi_i; i=1, ...,N\}$. The density $\rho({\bf r})$ and spin polarisation $m({\bf r})$ are given by
\begin{equation}
\rho({\bf r})=\sum_{i=1}^{N_{up}+N_{dn}} \mid \psi_i({\bf r})\mid^2,
\label{one}
\end{equation}
\begin{equation}
m({\bf r})=\sum_{i=1}^{N_{up}} \mid \psi_i({\bf r})\mid^2-\sum_{i=1+N_{up}}^{N_{up}+N_{dn}} \mid \psi_i({\bf r})\mid^2.
\label{two}
\end{equation}
The ground state energy and density are determined by minimising the KS energy functional
\begin{equation}
\begin{split}
E_{KS}[\rho]&=-\frac{1}{2}\sum_{i=1}^N\langle \psi_i\mid \nabla^2 \mid \psi_i\rangle+\int \rho({\bf r}) V_{ext}({\bf r}) d{\bf r}\\
&+\frac{1}{2}\int \int \frac{\rho({\bf r})\rho({\bf r'})}{\mid {\bf r-r'}\mid} d{\bf r} d{\bf r'}+U_{XC}[\rho],
\end{split}
\label{three}
\end{equation}
where $V_{ext}({\bf r})$ is the Coulomb potential of the positive charge distribution. Here $U_{XC}[\rho]$ is given by the local spin density approximation:
\begin{equation}
U_{XC}[\rho]=\int \rho({\bf r}) \epsilon_{XC}(\rho({\bf r}); \zeta({\bf r})) d{\bf r},
\end{equation}
where $\epsilon_{XC}(\rho({\bf r}); \zeta({\bf r}))$ is the exchange-correlation energy per electron~\cite{PZ} of the homogeneous electron gas
at the local density $\rho({\bf r})$ and local spin polarisation $\zeta({\bf r})=m({\bf r})/\rho({\bf r})$. 

Kohn-Sham orbitals are expanded on a basis of plane waves whose periodicity matches the cubic periodicity of the simulation cell:
\begin{equation}
\psi_i({\bf r})=\sum_{\bf G} c^{(i)}_{\bf G} e^{i {\bf G \cdot r}},
\end{equation}
where each ${\bf G}$ is a reciprocal lattice vector of the cubic simulation cell. The basis includes all plane waves whose ${\bf G}$ vector 
satisfies $G^2 \leq E_{cut}$ with $E_{cut}$ a suitable kinetic energy cut-off. A plane wave basis set of cut-off $2 E_{kin}$ is used to
represent the electron density and the external potential $V_{ext}({\bf r})$. The Fourier expansion of the latter is easily obtained by using
Poisson's equation, and considering that the Fourier transform of the positive charge density is given by
\begin{equation}
\tilde{\rho}({\bf G})=
\begin{cases}
\frac{2 \pi \rho_b}{V} L_z \frac{R_b^2}{2} & \text{if $G=0$,}\\
\frac{2 \pi \rho_b}{V} L_z \frac{(G R_b)}{G^2} J_1(GR_b)  & \text{otherwise,}
\end{cases}
\end{equation}
where $V=L^2 L_z$, and $J_1$ is the cylindrical Bessel function of order one. As implicit in our notation in Eq.~(\ref{one})-(\ref{three}), 
the Brillouin zone defined by the periodicity of the simulated system is sampled at the $\Gamma$ point only.
The size of the systems we simulate is such that this approximation does not introduce any sizable error. We also verified that the density overlap and the spurious
interactions across the $xy$ plane are negligible.

The optimisation of the $E_{KS}[\rho]$ functional is carried out by direct minimisation, i.e. by considering $E_{KS}[\rho]$ as an algebraic function of
the  Fourier coefficients $c^{(i)}_{\bf G}$, and using standard minimisation routines~\cite{hutter}. In doing so, we use the following expression for
the derivative of the Kohn-Sham energy functional with respect to the expansion coefficients $\{ c_{\bf G}^{(i)}\}$
\begin{equation}
\begin{split}
\frac{\partial E_{KS}}{\partial c_{\bf G}^{(j)\ast}}&=
\int \frac{\delta E_{KS}}{\delta \psi_j^{\ast}({\bf r})} \frac{d\psi_j^{\ast}({\bf r})}{d c_{\bf G}^{(j)\ast}}d{\bf r}=
\int \frac{\delta E_{KS}}{\delta \psi_j^{\ast}({\bf r})} e^{-i{\bf Gr}}d{\bf r}\\
& =\int \hat{H}_{KS}[\rho] \psi_j({\bf r}) e^{-i{\bf Gr}}d{\bf r}.
\end{split}
\label{seven}
\end{equation}
The last equality in Eq.~(\ref{seven}) implicitly defines the Kohn-Sham Hamiltonian $\hat{H}_{KS}$, which can also be written as
\begin{equation}
\hat{H}_{KS}= -\frac{1}{2} \nabla^2+V_{KS}({\bf r}),
\end{equation}
where $V_{KS}({\bf r})$ is the self-consistent KS potential.

\section{Results of the simulations}
\label{results}
\subsection{The ground state density distribution}
\label{ground}
Computations have been carried out for systems of $16$ to $32$ electrons, neutralised by a cylindrical background of aspect ratio $ 27.7 \leq L_z/R_b \leq 39.2$. 
Different total spin polarisations have been considered, from $N_{up}=N_{dn}=N/2$ to $N_{up}=N$, $N_{dn}=0$. Our computations, however, are
spin unrestricted, and even in the $N_{up}=N_{dn}$ case, spin polarisation can arise locally driven by exchange interactions.

We consider systems of fairly low density, corresponding to $25 \leq r_s \leq 40$. Previous studies on the homogeneous 3D electron gas, carried out using 
the same approach as ours, have shown that DFT-LSD predicts the onset of spin polarisation and charge localisation to take place at 
$r_s \sim 25$ ~\cite{giorgio, robin}, while electrons appear to be well localised at $r_s \geq 30$.

Our approach provides directly the ground state energy and density distribution for any given spin population as a function of size and shape of
the background density. The results for the total energy, however, display fairly predictable trends as a function of the background density and
linear density $\rho_{lin}$. For this reason, in full analogy with what is done in the classical case of atomic ions, we focus our discussion on the 
dependence of the density and spin distribution on the model parameters.

The linear to zig-zag transition in experimental quasi-1D ion systems is usually triggered by changing the geometric shape of the external potential
which confines the ions. As expected, stiff potentials that tightly confine ions in the radial direction, favour linear configurations, while
soft potentials give rise to zig-zag chains. In our model, the curvature of the external potential along the radial direction is directly related, via Poisson's 
equation, to the background density, and decreases with decreasing $\rho_b$. Therefore, we first present the results of computations for
samples of increasing $r_s$ (decreasing $\rho_b$), keeping fixed the number of electrons, the total spin, and the aspect ratio $L_z/R_b$.
We consider, at first, fully spin polarised systems, which arguably represent the simplest case.

\begin{figure}[t]
\includegraphics[scale=0.5]{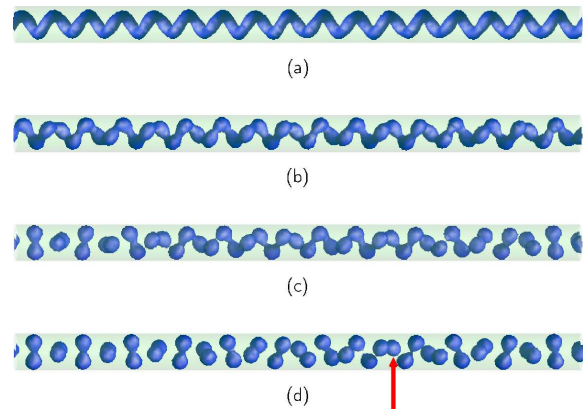}
\caption{(Color online)
Density iso-surface $\rho=2 \rho_b$ for samples of $N=32$ electrons, (a) $r_s=30$;  (b) $r_s=32$; (c) $r_s=35$; (d) $r_s=40$.
The vertical arrow marks the position of a defective, unpaired charge blob (see text).
}
\label{rhoofrs}
\end{figure}

The electron distribution is characterised by plotting density iso-surfaces, which in the homogeneous (3D) electron gas case show a regular pattern
of charge blobs for $r_s \geq 30$ (see Ref.~[\onlinecite{robin}]). The results for $N_{up}=N=32$, $N_{dn}=0$ at $r_s=30$, $32$, $35$, and $40$ are shown in 
Fig.~\ref{rhoofrs} (a) - (d). For all these systems the periodicity of the simulation cell along the wire axis is $L_z=32 r_s$, and the 
ratio $L_z/R_b$ is $27.7$. As already stated, the same periodicity is used along the other two directions, resulting in a large 
super-cell ($V=32768 r_s^3$), and a fairly high number of plane waves ($n_{pw}\sim 130000$) in the expansion of the KS orbitals.

The $r_s=30$ ground state density displays a clear helix geometry, with $17$ full periods within the simulation cell. The helix period is close but not 
equal to the length $\lambda=1/(2 k_F)$ expected on the basis of known singularities of the response function~\cite{twokf}, suggesting that at such low 
densities localisation cannot be quantitatively described in terms of linear response. In the equation above, $k_F$ is the Fermi wave 
vector in the $z$ direction, evaluated by computing KS bands in the 1D Brillouin zone, and assuming a cylindrically symmetric and translationally invariant 
charge density along the wire~\cite{zabala}. By comparing the number of particles and the number of helical turns in the simulation cell we see that slightly less than two 
electrons are accommodated in each turn. The non-integer number of electrons per turn suggests that the system might not be a 
closed-shell configuration. Therefore, the addition or the subtraction of electrons, while adjusting the background parameters to keep the system neutral, could 
enhance the ground state stability. We remark again that at $r_s=30$, DFT-LSD for the homogeneous electron gas gives a well localised ground state charge 
distribution at all spin polarisations. The helix configuration, therefore, results exclusively from the quasi-1D confining potential.

The discretisation of the charge density into localised blobs becomes apparent again already at $r_s=32$. Localisation, however, is still incomplete,
and the charge distribution gives rise to a crankshaft-like structure parallel to the wire axis. The crankshaft harms are marked by pairs of elongated and partially
overlapping blobs of electronic charge. According to the results of the minimisation process, the number of blobs in the simulation cell is $47$, i.e., 
significantly higher than the number of electrons in the system. 

\begin{figure}[t]
\includegraphics[scale=0.45]{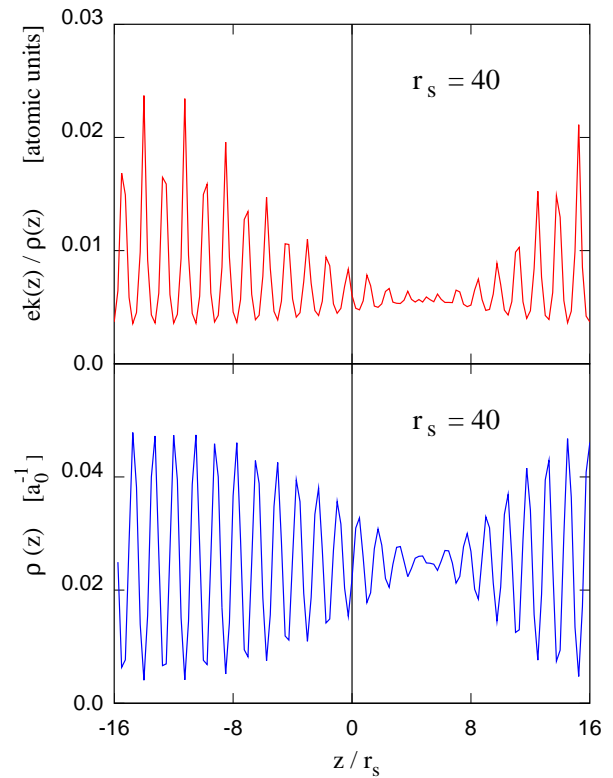}
\caption{(Color online)
Upper panel: planar average of the electron density $\rho(z)$ (see the definition in Sec.~\ref{results}).
Lower panel: planar average of the kinetic energy density $ek(z)$ divided by the corresponding density $\rho(z)$.
$ek(z)$ is  defined  in analogy with the planar density $\rho(z)$.
}
\label{rhoek}
\end{figure}

The partition of charge into blobs becomes progressively more marked with decreasing density, and is complete at $r_s=35$ (see Fig.~\ref{rhoofrs} (c)).
At $r_s=40$ (Fig.~\ref{rhoofrs} (d)) the charge distribution can be described as a line of charge dimers whose direction displays a rather complex pattern in space.
A closer analysis, however, reveals that this pattern can be described as due to a full rotation of the dimer bond about a direction perpendicular to the wire axis.
The dimer rotation is reflected in the $z$ dependence of the in-plane averaged density, defined as
\begin{equation}
\langle \rho(z)\rangle_{xy}=\int_0^L \int_0^L \rho({\bf r})dx dy
\label{1D}
\end{equation}
and shown in Fig.~\ref{rhoek} for $r_s=40$. Each of the density peaks in the rapidly oscillating part of this plot ($0 \leq z \leq 10 r_s$) marks the position of
dimers perpendicular (or nearly perpendicular) to the wire axis. The nearly constant portion at $z\sim 5 r_s$ corresponds to a few dimers 
nearly parallel to the $z$ axis. A similar behaviour is displayed by the in-plane average of the electron kinetic energy, shown in the upper panel of Fig.~\ref{rhoek} 
for a comparison.

The number of blobs remains constant at $47$ with decreasing density from $r_s=32$ to $r_s=40$, thus preventing the identification of blobs with single electrons. 
Clearly, the odd number of blobs implies that not all of them form dimers. One {\it defective} blob, in fact, is easily located and is identified by the vertical 
arrow in Fig.~\ref{rhoofrs} (d).
Apart from such an isolated defect, the structure of the fully polarised system at $r_s=40$ is fairly regular, with an intra-dimer 
separation almost exactly equal to $r_s$ and a dimer-dimer separation of $1.4 r_s$.

\begin{figure}[tbp]
\includegraphics[scale=0.55]{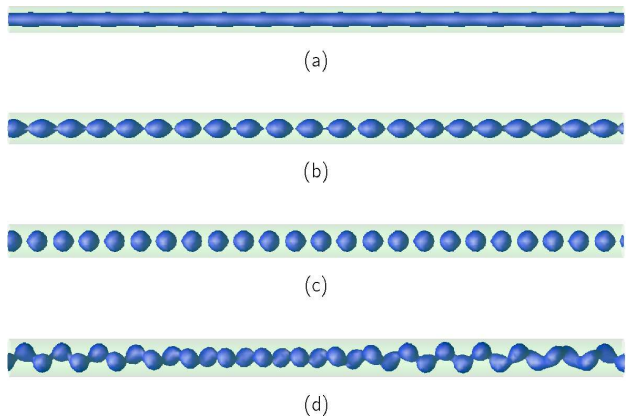}
\caption{(Color online)
Density iso-surface $\rho=2 \rho_b$ for fully spin polarised samples of equal length $L_z$ and different $R_b$ at $r_s=30$. 
(a) $N_{up}=N=16$; (b) $N_{up}=N=20$; (c) $N_{up}=N=24$; (d) $N_{up}=N=28$. 
}
\label{rhoofN}
\end{figure}

Our discussion has been focused, so far, on the dependence of the electronic structure on the curvature of the external potential. The shape of orbitals, however, 
is more directly related to the spatial extension of the self-consistent KS potential, which is very sensitive to the 1D density $\rho_{lin}$. To explore this dependence, 
we carried out computations for systems of the same length $L_z$, but different background radius $R_b$, corresponding to systems having a number of electrons between 
$16$ to $32$. The data for the density distribution at $r_s=30$ and full spin polarisation ($N_{up}=N$) are displayed in Fig.~\ref{rhoofN}, which shows that
with increasing $\rho_{lin}$ the distribution of charge blobs goes from linear towards more complex shapes, the transition taking place 
in between $N=N_{up}=24$ (linear) and $N=N_{up}=26$ (bent). 
More in detail, the $N=16$ and $N=18$ samples show a cylindrical charge distribution, $N=20$ to $24$ correspond to linear chains of blobs, elongated at first, and then
progressively rounded with increasing $N$. At $N=26$ the structure is a zig-zag chain, transforming into an helix above $N=28$. The sequence of ground state
structures as a function of $N$ is similar for $r_s=40$, with the exception that, at high $N$, the helix apparent in the $r_s=30$ data is replaced by a string 
of dimers at $r_s=40$. The transition from linear to non-linear configurations, in particular, takes place at the same size ($N=24$) at $r_s=30$ and $r_s=40$.

\begin{figure}[tbp]
\includegraphics[scale=0.48]{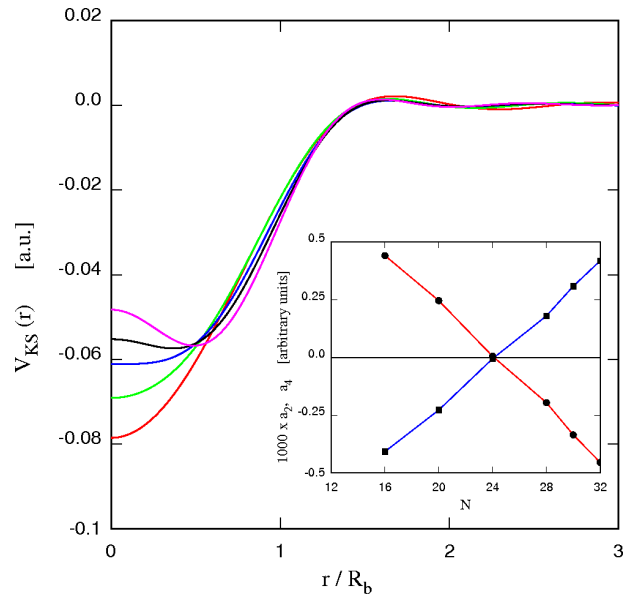}
\caption{(Color online)
Dependence of the Kohn-Sham potential on the radial distance $r$ for fully spin polarised samples of equal length $L_z$ and different $R_b$ at $r_s=30$.
(and thus of different values of the 1D density $\rho_{lin}$).
In order of increasing $V_{KS}(0)$, the curves correspond to: $N=16$, $N=20$, $N=24$, $N=28$, $N=32$. 
Inset: $N$-dependence of the quadratic ($a_2$) and quartic ($a_4$) coefficients in the polynomial fit of $V_{KS}(r)$ along the
radial direction. The $a_2$ coefficients have been rescaled (multiplied by $1000$) to plot $a_2$ and $a_4$ on the same scale.
}
\label{vks}
\end{figure}

As already suggested, the origin of these changes can be traced back to the Kohn-Sham potential, whose dependence on the radial distance is shown in Fig.~\ref{vks}.
The progressive widening and softening of $V_{KS}(r)$ with increasing $r_s$ is apparent in this figure and can be quantified by fitting $V_{KS}(r)$ at short $r$ with 
the sum of a quadratic and a quartic term. The fitting coefficients, given in the inset of Fig.~\ref{vks}, show that the quadratic term becomes rapidly less
important than the quartic one with increasing $N$, until it vanishes for $N\sim 24$. At the same time, the coefficient of the quartic term increases with increasing
$N$, thus preserving the overall stability of the system, at the expense of the linearity of the chain. 

Comparison of the charge distribution with the information given by $V_{KS}(r)$  suggests that the cylindrical charge distribution for $N=16-18$, as well as 
the elongated shape of blobs seen at $N=20$ is apparently due to the squeezing effects of a narrow harmonic potential. For $N> 24$, 
when the configuration is bent, confinement is exclusively due to the quartic term, while the negative quadratic term gives rise to an off-centre minimum. 
At the size of the transition ($N=24$), the short range portion of $V_{KS}(r)$ is very flat, and is well represented by a $r^6$ term. Further insight into the stability of the ground state structure found by our minimisations is obtained by changing the net spin of the sample. 

\begin{figure}[tbp]
\includegraphics[scale=0.55]{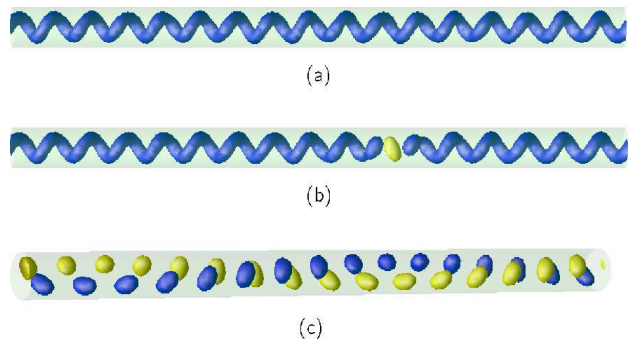}
\caption{(Color online)
Spin polarisation iso-surface $m({\bf r})=2 \ \rho_b$ at background density corresponding to $r_s=30$ for samples of $N=32$ electrons and
different spin populations. (a) $N_{up}=32$, $N_{dn}=0$; (b)  $N_{up}=31$, $N_{dn}=1$; (c)  $N_{up}=16$, $N_{dn}=16$.
Blue (dark) surface: spin up electrons. Yellow (light) surface: spin-down electrons.
}
\label{rhoofspin}
\end{figure}

Energy differences among configurations of the same size and $r_s$ but different $N_{up}$, $N_{dn}$ tend to be small at the densities considered in our
study. Nevertheless the results of our DFT-LSD computations shown that the spin-spin coupling turns from anti-ferromagnetic at $r_s=25$ to
ferromagnetic for $r_s \geq 32$. The evolution of the magnetic structure upon changing the ratio of $N_{up}$ and $N_{dn}$ at fixed background density is illustrated 
in Fig.~\ref{rhoofspin}, displaying magnetisation iso-surfaces for samples of $N=32$ electrons at $r_s=30$. Starting from the helical structure of the fully polarised 
case (Fig.~\ref{rhoofspin} (a)), reversing one spin in the $r_s=30$ case gives rise to a localised charge and spin blob that breaks the continuity of the helix
(see Fig.~\ref{rhoofspin} (b)). Even more striking and, at the same time, more significant, is the result obtained by reversing half of the $32$ electron spins. 
In the $r_s=30$ case, this breaks the single helix into a double helix, whose two strands have opposite spin (see Fig.~\ref{rhoofspin} (c)). 

\begin{figure}[b]
\includegraphics[scale=0.3]{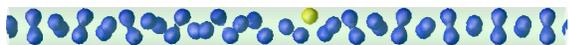}
\caption{(Color online)
Spin polarisation iso-surface $m({\bf r})=2 \rho_b$ at background density corresponding to $r_s=40$. System of $32$ electrons, $N_{up}=31$,
$N_{dn}=1$. 
Blue (dark) surface: spin up electrons. Yellow (light) surface: spin-down electron.
}
\label{oners40}
\end{figure}

\begin{figure}[b]
\includegraphics[scale=0.6]{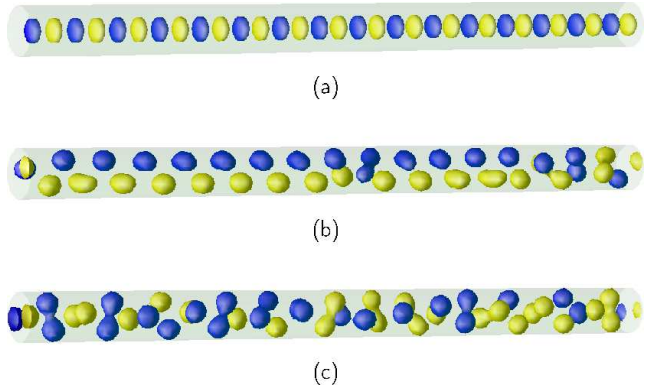}
\caption{(Color online)
Magnetisation iso-surfaces for samples of $N=32$ electrons, (a) $r_s=25$ ($\mid m \mid=0.3 \rho_b$);  (b) $r_s=35$ ($\mid m \mid=2 \rho_b$); (c) $r_s=40$ 
($\mid m \mid=2 \rho_b$). Blue (dark) surface: spin up electrons. Yellow (light) surface: spin-down electrons. The $r_s=30$ case is shown in Fig.~\ref{rhoofspin} (c). 
}
\label{mofrs}
\end{figure}

Reversing one spin on the $r_s=40$, fully polarised sample localises the reversed spin on the single blob not fitting into the dimer pattern described above (see 
Fig.~\ref{oners40}). This observation confirms our identification of this unpaired blob as a defect, whose stability is intrinsically lower than that of the dimerised 
blobs. Reversing now half of the spins in the $r_s=40$, $N=32$ sample results into a configuration somewhat similar to that of the $r_s=30$, $N_{up}=N_{dn}$ case. 
At this low density, however, a sizable amount of disorder makes the identification of the underlying double helix pattern more difficult. The enhancement of disorder 
is probably due to the further decrease of the exchange coupling with decreasing density. A comprehensive view of the dependence of density and magnetic
structures on $\rho_b$ (or equivalently $r_s$) in globally spin-compensated samples is given in Fig.~\ref{mofrs}.

\begin{figure}[tbp]
\includegraphics[scale=0.45]{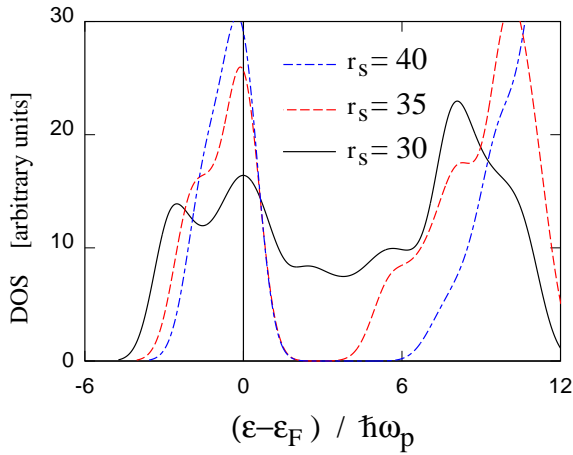}
\caption{(Color online)
Density of states for the Kohn Sham-eigenvalues of fully spin polarised samples at different background densities, $N=N_{up}=32$.
The Fermi energy is $\epsilon_F$, and $\omega_P$ is the plasmon frequency of the homogeneous electron gas at the same $r_s$,
given by $\omega_P=\sqrt{3}/r_s^{3/2}$.
}
\label{dos}
\end{figure}

The density of states computed from the Kohn-Sham eigenvalues in all cases consists of a few clearly identifiable bands, some of them overlapping. Especially
at low density, some of the bands are disjoint. The results for fully spin-polarised systems shown in Fig.~\ref{dos} confirms our anticipation that these samples 
are open-shell systems, whose Fermi energy falls in the middle of a band. The quantitative picture emerging from computations, however, depends on size and
spin, and in the case of the spin-compensated samples considered in our study ($N_{up}=N_{dn}=16$) the Fermi energy falls into a well defined and fairly wide 
gap for $r_s\geq 30$.

Conductivity has been computed using the Kubo-Greenwood formula, upon computing a fairly large number of empty states using the method briefly discussed in 
Ref.~[\onlinecite{hughes}]. The results (not shown) reflect the features seen in the DOS. The low frequency conductivity is highest for ferromagnetic samples, apparently 
because of their open-shell character. Moreover, in most cases we find non-negligible conductivity down to fairly low frequency, due to the presence of a few structural 
defects in the ground state electron distribution.

\subsection{Low-energy phonon-like excitations} 
\label{phons}

The subdivision of charge into weakly overlapping blobs motivates us to investigate the possibility of observing phonon-like excitations, corresponding to
small displacements of the centre of mass of individual blobs. This analysis, however, appears to be meaningful only for the cases where
the number of blobs corresponds exactly to the number of electrons in the system, in such a way that we can identify blobs with
single electrons. Our discussion below concerns one of such cases, corresponding to $N_{up}=N=24$, $r_s=40$. This system is in fact of particular
interest, since it marks the transition from linear to zig-zag configurations, and phonons are expected to reflect the impending change in the charge
distribution.

To estimate phonon frequencies, we approximate Kohn-Sham orbitals with single Gaussians, centred on the charge blobs or, more precisely, at
the position of the density maximum of each blob. That is, we consider:
\begin{equation}
\psi_i({\bf r})=A \exp{\{-\eta_1 [(x-X_i)^2+(y-Y_i)^2]-\eta_2 (z-Z_i)^2\}}
\end{equation}
where ${\bf R}_i\equiv (X_i, Y_i, Z_i)$ gives the position of blob $i$, initially set to coincide with the maximum of the corresponding electron density.
The set of orbitals is orthogonalised using the L{\"o}wdin algorithm~\cite{hutter}, which preserves the equivalence of all orbitals, and then is normalised.
The parameters $\eta_1$ and $\eta_2$ in the Gaussian exponent are varied in order to minimise the KS energy (at $r_s=40$ the optimal values of $\eta_1$ and $\eta_2$
are: $\eta_1 r_s^2=2.3552$, $\eta_2 r_s^2=3.52$). The approximation is remarkably accurate at  $r_s=40$, as confirmed by the low increase (less than $3 \times 10^{-5}$ 
Ha per electron) of the optimal energy with respect to the plane wave estimate, based on the unconstrained optimisation of $\sim 130000$ coefficients per orbital.

\begin{figure}[tbp]
\includegraphics[scale=0.45]{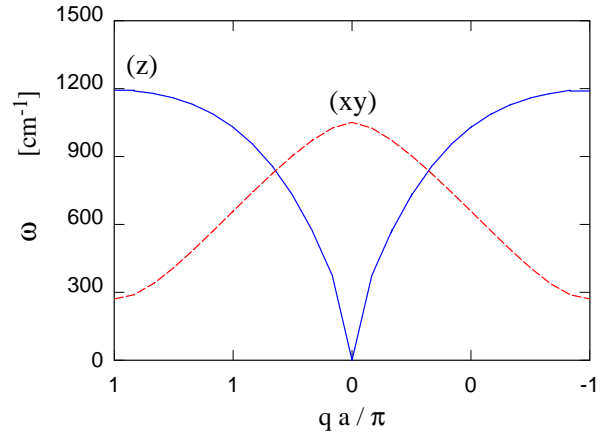}
\caption{(Color online)
Phonon dispersion relation for a fully spin polarised sample of $N=24$ electrons at $r_s=40$. The 1D Brillouin zone (along $z$) refers to the minimal
unit cell of lattice constant $a=L_z/N$, containing one single electron.
}
\label{phon}
\end{figure}

The Hessian $\partial^2 E_{KS}/\partial R_I^{\alpha} \partial R_J^{\beta}$ for the $N=24$ supercell is computed by numerical differentiation 
of the energy upon moving the centre of Gaussians $I$ and $J$ by a small displacement $\delta$ along the coordinates $\alpha$ and $\beta$, respectively. The resulting matrix is combined with a kinetic part, to give the dynamical matrix of the chain, whose diagonalisation
provides an estimate for the phonon frequencies. Analysis of the eigenvectors, or, more precisely, of the $z$-dependence of their polarisation vector,
allows us to associate each eigenfrequency to a momentum $q_z$, and thus to draw the dispersion relation over the first Brillouin zone.
The results are shown in Fig.~\ref{phon}. Phonon frequencies belong to three branches, two of them being degenerate. The non-degenerate branch
corresponds to vibrations along the $z$ direction. It is an acoustic branch, whose frequency vanishes at $q_z=0$, then increases monotonically
in moving towards the zone boundaries. The two other branches correspond to vibrations in the $xy$ plane. The frequency of their $q_z=0$ modes does not vanish, 
because of the restoring force due to the external potential. Frequency, however, decreases with increasing $\vert q_z\vert$, nearly vanishing
at the zone boundary. This behaviour clearly points to the easy deformation of the linear chain obtained by displacing charge blobs perpendicularly
to the wire axis, with nearest neighbouring atoms moving into opposite directions. The nearly soft mode thus corresponds to the formation of 
transversal dimers. 

This result suggests that the linear to non-linear transition is second order, or at most weakly first order, in qualitative agreement with the findings of
Ref.~[\onlinecite{fishman}] for the analogous transition in classical ion systems.

\section{Summary and conclusions}
\label{summary}

Low dimensional assemblies of ions trapped into static and time-dependent electromagnetic fields have been extensively investigated in the past, both
computationally and experimentally, and could find applications in metrology and in quantum information. At the conditions of
present experiments, the motion of ions is classical, even though quantum mechanical aspects (strictly required for quantum information applications) are
associated to the orientation and dynamics of spin moments.

Recently, many experiments and calculations have been devoted, in particular, to analyse the so-called linear to zig-zag transition~\cite{report} 
taking place  in nearly 1D ion systems upon varying the linear density of ions and/or changing the geometric parameters of the confining
potential. In our paper, we have investigated the geometric and magnetic structure, electronic properties and low energy phonon-like excitations of nearly 1D electron
systems confined by the electrostatic potential of a very thin cylindrical distribution of positive charge. We have considered a wide range of
relative spin populations $N_{up}$ and $N_{dn}$ and focused on the low density regime ($25 \leq r_s \leq 40$), where
electrons tend to localise giving rise to blobs of negative charge distributed in space. In many respects, this model represents the 
quantum counterpart of the classical ion systems.

The results of our computations, carried within DFT-LSD, show that the quantum system displays a much wider variety of configurations and properties than 
in the classical case. In fully spin-polarised systems, we observe the stability of helicoidal density distributions for $ 25 \leq r_s \leq 30$, turning 
into a twisted string of localised charge dimers at lower density. Spin-compensated samples at $ 25 \leq r_s \leq 30$ display an intriguing double-helix structure,
whose two strands have opposite spin polarisation. The double helix unrolls into two linear chains of opposite spin at densities around $r_s=35$, providing one of the
few examples of virtually planar, zig-zag configuration found in our computations. 
Also in the case of globally spin-compensated samples, charge dimers form at lower density ($r_s > 35$) with a predominantly ferromagnetic coupling within 
each dimer and anti-ferromagnetic coupling among dimers. The spin-spin coupling, however, is low at densities such that $r_s \geq 30$, and a sizable amount 
of disorder is observed in the distribution of spins.  At all densities, computations with a single spin-reversed impurity in an otherwise ferromagnetic sample 
reveal localised magnetic and structural defects, which might dominate the response of low density electron chains to external perturbations.

The density of states for the Kohn-Sham eigenvalues consists of several bands, some of them partially overlapping, some other disjoint, according 
to density, aspect ratio and spin population. The ferromagnetic samples analysed in our study tend to be open-shell systems up to the lowest densities
we investigated ($r_s=40$), while spin-compensated samples are closed shell systems with a fairly wide gap separating occupied KS states from
unoccupied ones. The open-shell character of the ferromagnetic systems, together with the defects found in fully or partially spin-compensated cases, 
give rise to the non negligible low-frequency conductivity predicted by the Kubo-Greenwood formula for most of the samples investigated in our study.

To the best of our knowledge, linear assemblies of electrons have never been made experimentally by the techniques used to trap atomic ions, although 
interesting experimental efforts on the single electron scenario are paving the way to an up-scaling. Our results provide additional motivations for experimental 
investigations along this line. 
Our findings suggest that electron systems of this kind, if ever realised, would exhibit a broad range of unusual and surprising properties, which
could also find useful applications. 

From a fundamental standpoint, our analysis provides a useful complement to theoretical results obtained using classical simulation approaches by emphasising
quantum mechanical and spin effects that might become important even in ion systems at sufficiently
low $T$. Moreover, the density iso-surfaces computed in our study provide a more comprehensive view of the ground state properties than the information given by
low energy geometries of classical many-particle systems. The elongated structures seen in some of the density iso-surfaces describe above, for instance,
are a pictorial representation of low energy valleys in the potential energy surface, suggesting patterns for low frequency excitations.

It is worth remarking that many-electron systems of this kind are already being made in condensed matter, using controlled doping to introduce mobile 
electrons into polymers and into semiconductor nanostructures~\cite{matti}, giving rise to quasi-1D structures whose size and density are not far from those considered in 
our study.  In these cases, however, the measurement of the properties of the system and their interpretation are certainly more difficult than in the ion-trap scenario.

Our work paves the way to a variety of new studies. For instance, an intriguing point to tackle would be the control of one of  the ``defects'' found in our simulation 
samples. This could be realised by studying, via time-dependent density functional theory, the dynamics of the defect and its motion across the chain as induced by an 
external perturbative potential. It would be very interesting to study the effectiveness of this scenario for quantum communication protocols such as quantum state 
transfer, which are based on the use of quasi-1D lattices of interacting particles.

\acknowledgments

We thank G. De Chiara and G. Morigi for helpful discussions. MP acknowledges financial support from the UK EPSRC (EP/G004579/1).

\end{document}